# Adiabatic reversible compression: a molecular view

E. N. Miranda

CRICYT - CONICET

5500 - Mendoza, Argentina

and

Dpto. de Física

Universidad Nacional de San Luis

5700 - San Luis, Argentina

**Abstract:** The adiabatic compression (or expansion) of an ideal gas has been analysed. Using the kinetic theory of gases the usual relation between temperature and volume is obtained, while textbooks follow a thermodynamic approach. In this way we show once again the agreement between a macroscopic view (thermodynamics) and a microscopic one (kinetic theory).

The student that has a first contact with thermodynamics generally becomes aware also with the kinetic theory of gases [1, 2, 3]. The microscopic view is used to obtain the state equation for perfect gases, but the microscopic approach is set aside in favour of the thermodynamic one. The aim of this article is to analyse the adiabatic compression from the kinetic point of view. In this way, students have another simple example where the microscopic and macroscopic approaches converge. The molecular view enlightens the compression (or expansion) process because the student can visualise the collisions between molecules and the piston and the consequent exchange of kinetic energy. Additionally, a classical analysis of the degrees of freedom of the molecules explains the origin of the exponent in the usual formula for an adiabatic compression or expansion.

This article is divided into two sections. In the first one is sketched how to obtain the relevant formula for an adiabatic compression or expansion using the thermodynamical approach. It is included just for completeness and can be skipped by the reader. In the second section the mentioned equation is attained from a molecular point of view.

**I - Thermodynamical approach**

In this section, we derive the expression for an adiabatic reversible compression or expansion of one mole of an ideal gas by thermodynamic considerations. There are several ways for doing this, and the most direct one has been chosen. Symbols are used in their standard meanings as presented in well-kown textbooks [1-5].

Let us imagine a situation with a cylinder and a piston as shown in Fig. 1. The piston moves very slowly (i.e. quasi-statically) toward the left. The volume changes from $V^i$ to $V^f$. The walls are adiabatic, i.e. there is no heat exchange with the surroundings; consequently $\delta Q = 0$. Furthermore, the only work involved is that of compression, hence $\delta W = -p.dV$. Replacing into the expression of the first principle, we get:

$$dU = \delta Q + \delta W = -p\,dV \tag{1}$$

Since we are dealing with an ideal gas, the internal energy depends only on the temperature $T$ through the relation $dU = C_v\, dT$. The law of ideal gases allows us to write $p = RT/V$; By use of eq. (1) we obtain:

$$C_V\, dT = -RT \frac{dV}{V}$$
$$\frac{dT}{T} = -\frac{R}{C_V} \frac{dV}{V} \qquad (2)$$
$$\frac{dT}{T} = -(\gamma - 1) \frac{dV}{V}$$

To get the last line in eq. (2), one should remember that for ideal gases $R = C_p - C_v$, and that $\gamma$ is the ratio between the specific heats $\gamma = C_p/C_v$. For ideal monoatomic gases it is $\gamma = 5/3$ and for diatomic molecules $\gamma = 7/5$. A simple integration of (2) between the initial conditions $(T^i, V^i)$ and the final ones $(T^f, V^f)$ leads to the final result:

$$\ln\left(\frac{T^f}{T^i}\right) = -(\gamma - 1) \ln\left(\frac{V^f}{V^i}\right)$$
$$\frac{T^f}{T^i} = \left(\frac{V^i}{V^f}\right)^{\gamma - 1} \qquad (3)$$

The last formula can be rewritten by using the ideal gases law as the common expression $p^i (V^i)^\gamma = p^f (V^f)^\gamma$. However (3) is better for our purpose. Our aim is to attain (3) using the kinetic theory of gases.

**II - Molecular approach**

Let us consider the situation shown in Fig. 2: a molecule of mass $m$ moves with an horizontal speed component to the right $u_x$ and collides with a piston of mass $M$ that is moving toward the left with velocity $w$. The walls of the cylinder are adiabatic which, at the molecular level, means elastic collision. Consequently, from energy conservation we can write:

$$\tfrac{1}{2} m(u_x)^2 + \tfrac{1}{2} Mw^2 = \tfrac{1}{2} m(u'_x)^2 + \tfrac{1}{2} M(w')^2 \qquad (4)$$

$u'_x$ y $w'$ are respectively molecular horizontal speed component and piston speed after the collision.

On the other hand, linear momentum conservation leads us to:

$$mu_x + Mw = mu'_x + Mw' \qquad (5)$$

From (4) and (5):

$$u'_x = \left[-u_x\left(1-\frac{m}{M}\right)+2w\right]\left(1+\frac{m}{M}\right)^{-1} \tag{6a}$$

and:

$$w' = \left[2u_x\frac{m}{M}+w\left(1-\frac{m}{M}\right)\right]\left(1+\frac{m}{M}\right)^{-1}$$

(6b)   Equations (6) can be simplified since $\frac{m}{M} \cong 0$ (and $\frac{w}{u_x} \cong 0$); therefore:

$$u'_x = -u_x + 2w$$

(7a)

$$w' = w \qquad \text{(quasi-static process)} \tag{7b}$$

Notice that:

$$(u'_x)^2 \cong (u_x)^2 - 4wu_x \tag{8}$$

Our next step is to relate the result of eqs. (7) to a mean molecule of one mole of an ideal gas which moves with the root mean speed $c$ when in thermoynamics equilibrium at temperature $T$. We can write:

$$\begin{aligned}c^2 &= u_x^2 + u_y^2 + u_z^2 \\ c^2 &= 3u_x^2\end{aligned} \tag{9}$$

and

$$\varepsilon = \frac{1}{2}mc^2 = \frac{3}{2}kT \tag{10}$$

by the theorem of equipartition of energy, $k$ being the Boltzmann constant and $\varepsilon$ the mean kinetic energy of the molecule.

Now, let us determine the number of collisions of all molecules with the piston per unit time, $N_c$, for a given position $x$ of the piston. A bit of reflection shows that:

$$N_c = (N_A . u_x)/2x \tag{11}$$

$N_A$ is the Avogadro number -i.e. the total number of molecules in one mole-. To get this result consider a time interval $\Delta t$; the number of collision with the piston is $u_x.\Delta t.A. \rho/2$ where $A$ is the piston area and $\rho$ is number of molecules per unit volume, i.e. $\rho = N_A / A.x$. The factor 2 appears because only half of the molecules move toward the piston. Since we are interested in the number of collisions per unit time, equation (11) is obtained.

At each impact there is a small change in the value of $(u_x)^2$ given by eq. (8).

Consequently, the change in the kinetic energy $\varepsilon_x$ associated with the motion in the $x$ direction is given by:

$$\Delta\varepsilon_x = \Delta(\tfrac{1}{2}mu_x^2) \\ = -2mu_x w \tag{12}$$

We are able to write the change per unit time in the total kinetic energy $E_x$ due to the motion in the $x$ direction. It will be given by the number of collisions per unit time multiplied by the energy change in each collision:

$$\frac{dE_x}{dt} = N_c \Delta\varepsilon_x = -\frac{2w}{x}E_x \tag{13}$$

Since the only cause in the total energy $E$ is due to variations in the energy associated with motion in the $x$ direction, we may write:

$$dE = dE_x \tag{14}$$

We are dealing with a monoatomic gas without internal degree of freedom. For that reason, the theorem of equipartition of energy allows to write:

$$E = 3E_x \tag{15}$$

Taken into account the linear relation between energy and temperature – eq. (10) – and eqs. (13), (14) and (15), it may be written:

$$\frac{dE}{E} = \frac{dT}{T} = -\frac{2}{3}\frac{w}{x}dt \\ = -\frac{2}{3}\frac{dx}{x} \tag{16}$$

Since $V(x) = A.x$, we may write $dx/x = dV(x)/V(x)$. In this way, the expression is ready for integration between the initial ($V^i$, $E^i$) and final ($V^f$, $E^f$) conditions:

$$\frac{E^f}{E^i} = \left(\frac{V^i}{V^f}\right)^{2/3} \tag{17}$$

$$\frac{T^f}{T^i} = \left(\frac{V^i}{V^f}\right)^{2/3} \tag{18}$$

Eq. (17) is identical to (3) for the case of ideal monoatomic gases ($\gamma = 5/3$).

The exponent comes from the fraction of the total energy due to motion in the $x$-direction. In a diatomic molecule there are 5 degrees of freedom ( 3 due to translation and 2 to rotation). Then, the theorem of equipartition of energy reads:

$$E_{tot} = 5E_x \tag{19}$$

The mathematical steps are the same as in the previous case, and we finally get

$$\frac{T^f}{T^i} = \left(\frac{V^i}{V^f}\right)^{2/5} \tag{20}$$

The origin of the exponent in the usual formula is clear now: it is intimately related to the equipartition of energy.

In summary, the aim of this paper has been to give an example simple enough to show how the kinetic theory of gases and thermodynamics lead to the same result for the same problem. The level is suitable for a student with a general background in physics and who is having his first contact with thermodynamics and kinetic theory. In section I, the reversible adiabatic compression of an ideal gas was studied from a thermodynamical point of view. The formula shown in (3) -a standard textbook result- is obtained. In section II the same problem has been considered from the molecular stand-point . A straightforward application of energy and linear momentum conservation allows evaluating the change in the molecular kinetic energy due to a collision -eq. (12)-. The theorem of equipartition of energy is central in the analysis: one should know the fraction of the total energy due to the motion in the compression direction -eqs. (15) and (19)-. From this point and using simple mathematics, one gets eq. (18) -or (20)- that agrees with the formula obtained with the thermodynamical approach; and the aim of the work has been accomplished.

**Acknowledgement**: The author is supported by the National Research Council (CONICET) of Argentina. The author thanks the unknown referee for a careful reading of the manuscript and many suggestions.

**Figure captions**:

**Figure 1**:

This figure shows the general setting used for the calculations. A piston is initially at $L^i$ and it moves slowly (quasi-statically) to the final position $L^f$. In the figure, the piston is shown in a generic position $x$. The piston area is $A$, the initial volume is $V^i$ and the final one $V^f$. For the generic position shown, the volume is $V(x)$. Notice that all the walls are adiabatic.

**Figure 2:**

A molecule of mass $m$ moves toward the right with speed $u_x$ and collides with the piston of mass $M$ that moves rightward with velocity $w$. After the collision, the molecule speed is $u'_x$.

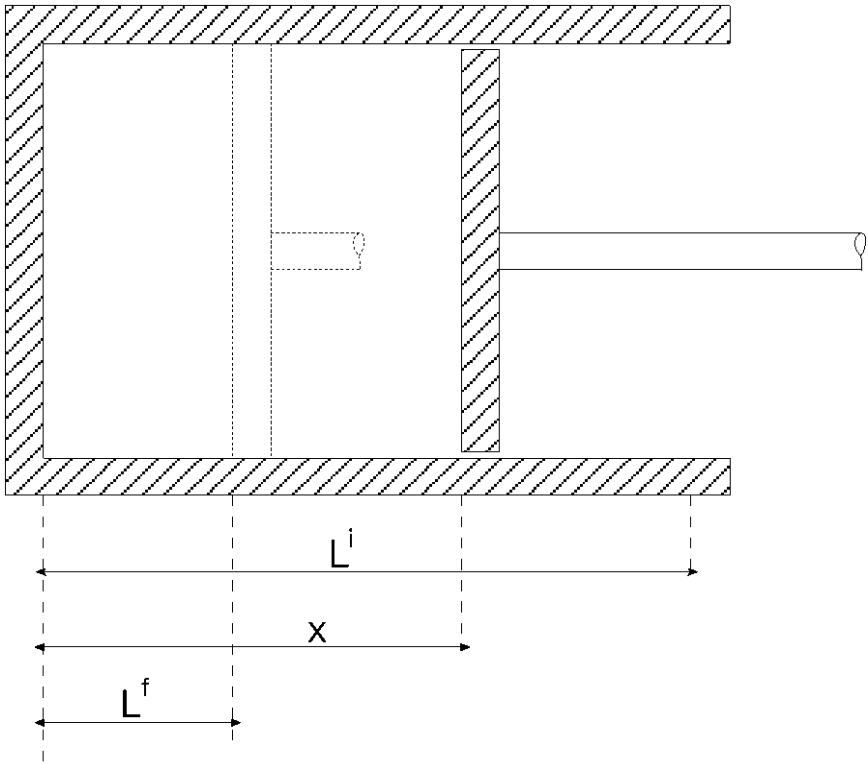

Figure 1

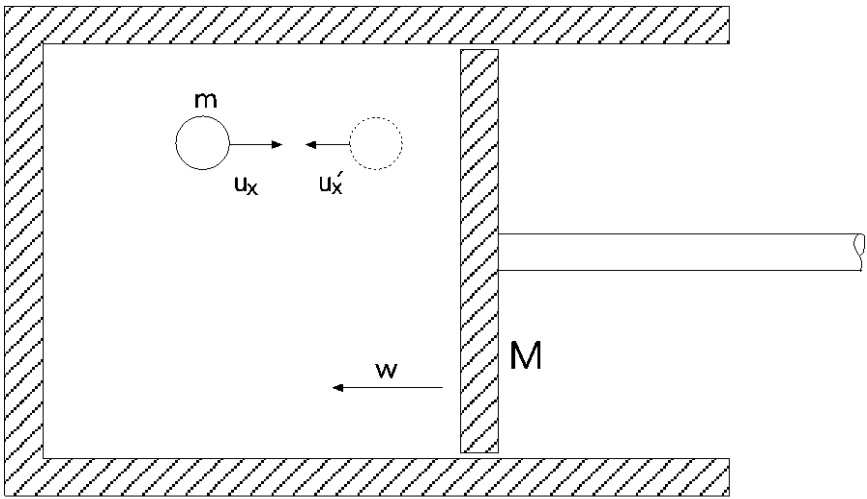

Figure 2